\documentclass[]{spie}  

\usepackage{amsmath,amsfonts,amssymb,decorule}
\usepackage{graphicx}
\graphicspath{{images/}}
\usepackage[colorlinks=true, allcolors=blue]{hyperref}
\usepackage{booktabs, microtype, siunitx}
\usepackage[skip=1pt]{caption}

\title{Infant Brain Age Classification: 2D CNN Outperforms 3D CNN in Small Dataset }

\author[a]{Mahdieh Shabanian} 
\author[b]{Markus Wenzel}
\author[c]{John P. DeVincenzo}

\affil[a]{Department of Biomedical Engineering, University of Tennessee Health Science Center, Memphis, TN, United States}
\affil[b]{Fraunhofer Institute for Digital Medicine MEVIS, Bremen, Germany}
\affil[c] {Enanta Pharmaceuticals, Watertown, MA, United States}

\authorinfo{Further author information: Send correspondence to Mahdieh Shabanian\\ E-mail: shabanian.m@gmail.com}

\begin{document} 
\maketitle

\begin{abstract}

Determining if the brain is developing normally is a key component of pediatric neuroradiology and neurology. Brain magnetic resonance imaging (MRI) of infants demonstrates a specific pattern of development beyond simply myelination. While radiologists have used myelination patterns, brain morphology and size characteristics to determine age-adequate brain maturity, this requires years of experience in pediatric neuroradiology. With no standardized criteria, visual estimation of the structural maturity of the brain from MRI before three years of age remains dominated by inter-observer and intra-observer variability. A more objective estimation of brain developmental age could help physicians identify many neurodevelopmental conditions and diseases earlier and more reliably. Such data, however, is naturally hard to obtain, and the observer ground truth not much of a gold standard due to subjectivity of assessment. In this light, we explore the general feasibility to tackle this task, and the utility of different approaches, including two- and three-dimensional convolutional neural networks (CNN) that were trained on a fusion of T1-weighted, T2-weighted, and proton density (PD) weighted sequences from 84 individual subjects divided into four age groups from birth to 3 years of age. In the best performing approach, we achieved an accuracy of 0.90 [95\% CI:0.86-0.94] using a 2D CNN on a central axial thick slab. We discuss the comparison to 3D networks and show how the performance compares to the use of only one sequence (T1w). In conclusion, despite the theoretical superiority of 3D CNN approaches, in limited-data situations, such approaches are inferior to simpler architectures. The code can be found in \url{https://github.com/shabanian2018/Age_MRI-Classification}
\end{abstract}

\keywords{Deep learning, CNN, neurodevelopmental, age estimation, infant diseases.}

\section{INTRODUCTION}
\label{sec:intro}  

Neurodevelopmental disorders (NDDs) are a diverse group of conditions characterized by delayed milestones involving cognition, communication, behavior, and motor skills.  Infancy and early childhood are characterized by rapid cognitive development, especially in the first three years of life. This cognitive development is mirrored by changing brain structure, function, and connectivity. Many pediatric diseases impair this development. Therefore, brain developmental age estimation is crucial to determine if a child's brain is developing normally. Magnetic resonance imaging (MRI) based neuroimaging of infants offers qualitative information such as myelination patterns and brain morphology. MRI also offers quantitative information such as head circumference, brain volume, and water content. Neuroimaging is critical for determining the impact of pediatric brain diseases, in particular when compared to a healthy clinical sample of normal brain development, to help with many infant diseases, including prematurity \cite{van2019prevention}, hypoxic ischemic encephalopathy (HIE) of the newborn \cite{shetty2019cerebral}, congenital cytomegalovirus (cCMV) infection \cite{grinberg2019volumetric}, bacterial meningitis, herpes simplex virus encephalitis (HSVE) \cite{ramirez2018comparing}, pediatric epilepsy, cerebral palsy (CP), tuberous sclerosis complex (TSC) and many other genetic and non-genetic disorders. 

Newborns with brain disorders are at increased risk of severe long-term motor deficit and cognitive delay. Detecting these conditions early could help mitigate morbidity. The developing brain's distinct morphological changes are the basis of image-based brain age determination. In the light of observer variability and the complex information for humans to assess, computer-based automated brain developmental age estimation (BDAE) could offer an important automated second opinion in patients with mild cognitive impairment or a tool to monitor neurodevelopment in patients with known chronic neurologic diseases. 

In infants, both imaging and computational approaches have yet to achieve an acceptable level of consensus. In the first year of life, myelination occurs very rapidly with gray matter migrating from its origin in the periventricular region to the cortex. These properties change rapidly thereafter and lead to many computational challenges not seen in adult models. As a result, neonatal brain development, as observed by MRI, can be roughly divided into four distinct temporal stages \cite{paus2001maturation}. (1) Newborn pattern; myelination only present along the posterior limbs of the internal capsule and perirolandic regions.  The major sulci and gyri are well developed. Myelination involves the genu of the corpus callosum, although a majority of the T1w and T2w signal remains switched around 3 months. The majority of the white matter is hyperintense in T2w, making it difficult to differentiate from cerebral edema. (2) Characterized by further progression of myelination anteriorly (5-9 months of age), superiorly, and laterally. The splenium of the corpus callosum becomes myelinated during this time. Around 12 months, the rate of myelination drops off, with tissue contrast more closely resembling a fully developed brain on T1-weighted imaging. (3) After 2 years, essentially the adult myelination pattern seen on T1-weighted imaging, however T2-weighted sequences will still have areas of bright signal that are not found in an adult. (4) T2 hyperintense signals consistent with white matter tracts that have yet to be myelinated can be found in the young adult, especially in the frontal lobes.    

With no biomarkers to diagnose NDD or quantitative assessment of neurological pathology in infants, these NDDs are identiﬁed through physical exams and imaging tests. Then improvements in the identification of MRI abnormalities is needed in infants to predict long term damage and to assign earlier interventions. Automated BDAE algorithms applied to infants who were imaged with MRI in clinical care could thus provide a reproducible, automated second opinion to the clinicians’ assessment. Current pediatric neuroradiology assesses the age of the patient predominantly from myelination of the brain and brain volume. Such an approach suffers from interobserver and intraobserver variability. Artificial intelligence algorithms may make it possible to evaluate early signs of delayed age more objectively and quantitatively for each patient. Deep learning (DL) is one of the most successful variants of machine learning, utilizing deep artificial neural networks (ANNs). These hierarchically organized ANNs are able to learn efficient feature representations of input data \cite{mostapha2019role,jahangard2020u,cui2020introduction, amagasakibrain}. State-of-the-art BDAE approaches use manually extracted features, preventing subsequent machine learning approaches from fully exploiting the MR images’ contents \cite{wang2014age,franke2010estimating,lao2004morphological}. 

We hypothesize that a custom 3D deep learning approach (a 3D Convolutional Neural Network, 3D CNN) could be used to differentiate age-appropriate brain development from retarded development based on multi-modal MRI sequences. As it will be shown, this is not the case, and we examine possible reasons by providing different types of reduced information to the DL approach: 3D data, 2.5D data, and 2D data. The ultimate aim of this work is to approach a continuous (rather than age-group stratified) estimation of the developmental brain age from MRI. But even a classification into age groups will help to identify patients with developmental delay and serve as an objective finding that could help eliminate inter and intra-observer variability. It could also be used to risk-stratify different patient populations.

\section{Materials and Methods}
\subsection{Data}

We obtained 552 normal MRI scans (equaling 184 sets of T1w, T2w, and PDw sequences) of 84 infants from the NIMH Data Archive (NDA) from 8 days to 3 years, spanning a critical developmental period to enable early diagnosis of neurological sequelae. Some included infants had several exams at different ages. In all training-validation splits, we ensured that these patients were always assigned to training or validation sets only. The patients ranged in age from 8 days to 3 years, spanning a critical developmental period to enable early diagnosis of neurological sequelae. The infants were scanned with 1.5T MRI while awake or during natural sleep without sedation. The infants were scanned with 1.5T MRI while awake or during natural sleep without sedation. MRI acquisition generally lasted 30–45 minutes on a 1.5T scanner with a 2D sequence that minimized scan duration for the newborn to 3 years group. The axial scans consisted of a 2D T1-weighted (T1W) spin echo and a T2-weighted (T2W) 2D turbo spin-echo sequence. The T1W sequences utilized repetition time (TR) 500, echo time (TE) 12, 90-degree flip angle, and 3-mm slice thickness.  The T2W and PD-weighted (PDW) scans utilized TR 3500, TE 15-17 (115-119), and 3-mm slice thickness.  The T1W and T2W scans were nominally 1×1×3 resolution (1×1×3 or .97×.97×3) with a matrix of 256 x 192 mm. Most scans were obtained using a Sonata, Magnetom scanner (Siemens Medical Systems, Erlangen, Germany), while another site used a GE Signa Excite scanner (GE Healthcare, Chicago, IL) to obtain less than half of the scans.  \cite{sanchez2012neurodevelopmental}. 

This dataset was divided into four nominal age cohorts that may be useful for clinical applications but not for designating typical developmental patterns. The cohorts we chose were: Newborn (n=47), 1-year-old (n=60), 2-year-old (n=26), and 3-year-old (n=51). Note that the actual age of the subject will deviate from this classification, particularly for subjects with TSC. Each patient's data is comprised of T1-w, T2-w, and PD-w sequences acquired either on a Siemens or a GE scanner, depending on the site of clinical origin. Images in this study were provided as Neuroimaging Informatics Technology Initiative (NIfTI) image volume datasets. The NIMH data was accessed under the terms of the NDA.

\subsection{Model Details}
For our estimation of brain developmental age, we designed custom 2D, 2.5D (not reported here) and 3D fusion CNN architectures. We slightly downsampled all MRIs to 250x250x40 voxels and we cropped a center region to 150×150×20 voxels for 3D model and to 150×150 for 2D model to reduce computational complexity while maximizing the amount of information retained from the original resolution. Our 3D proposed architecture has four scale level blocks consisting of 1) 2-2-2-1 (3x3x3) convolutional layers followed by 2) 2x2x2 max pooling layers,and a global average pooling layer before 3) three fully-connected (FC) layers that lead into 4) the SoftMax four-class output layer. The convolutional blocks feature 32, 64, 96, and 128 channels, respectively. We used two blocks of 2x2 convolutional layers, four 64 channels and two 128 channels for our 2D proposed model with 2x2 max pooling and three FC layers as a simple model. Following best practices, cross-entropy divergence loss and the Adam optimization algorithm (lr=0.001) were used for the loss function and optimizer, respectively in both the 2D (368,580 parameters) and the 3D model(392,516 parameters). The model weights were randomly initialized. Since no separate test data was available, we used a stratified 5-fold cross-validation scheme to train and validate all CNN models. In 84 patients with partially multiple visits, this resulted in 139-154 images in the training set and 30-45 images in the validation set. We performed data augmentation to reduce potential overfitting and monitored training and validation loss accordingly. Each image in the training set of both model was augmented by L/R ﬂipping, static rotations around the z axis by $\pm$15 degree, and random rotations of $\pm$15-65 degree. All our models were trained on a personal computer (NVIDIA TITAN RTX GPU, Python 3.7.9, TensorFlow 2.1.0).

\section{Experimental Evaluations}
To measure the performance of classification of MRIs in each category, we calculated six metrics in the entire validation dataset: positive predictive value (PPV), true positive/negative rate (TPR, TNR), F1-score (FS), accuracy (ACC), area under the receiver operating characteristic curve (AUROC) and Matthew's Correlation Coefficient (MCC). These metrics are useful to characterize the performance of models in multi-classification tasks on imbalanced datasets, where accuracy is a misleading metric if reported alone~\cite{vidiyala_ramya_performance_2020}. We reported the metrics according to the four classes of brain developmental age and the classification errors in a normalized confusion matrix for the most accessible visualization.

\section{Results}

\subsection{Baseline Experiment}
Several studies used only T1-weighted MRI volumes to estimate brain age using CNN \cite{cole2017predicting,jnawali2018deep,ueda2019age}. Therefore, we initially performed a corresponding experiment using only the T1w scans in our proposed CNN models. We compared them through 5-fold cross-validation similar to the methodology used when evaluating the fusion models. Dropout (0.7) and (0.2) used respectively in first and second FC layers and batch normalization  further help to improve model convergence.

\begin{figure}
\centering
    \includegraphics[width=\linewidth]{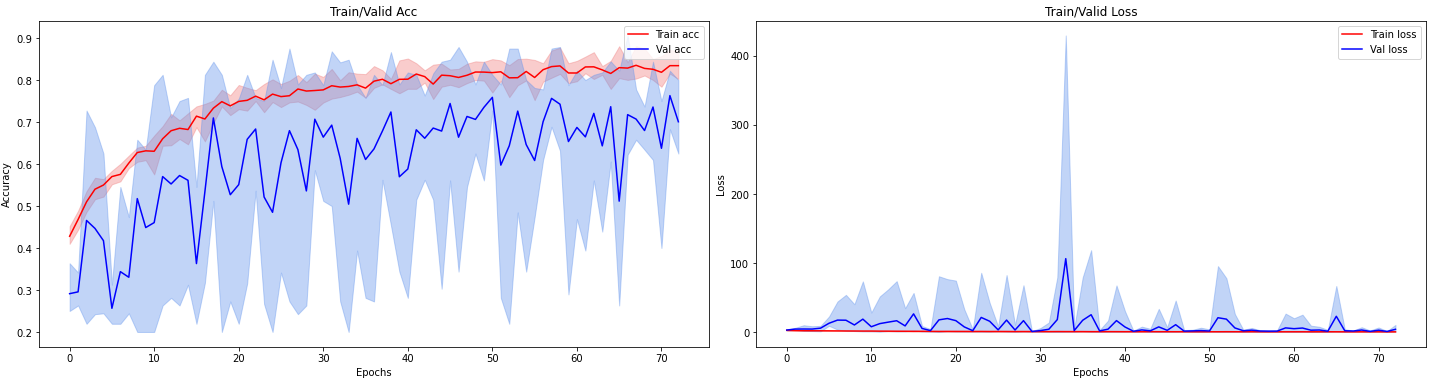}
    \caption[Accuracy/loss in 2D CNN in T1w, averaged for  5  folds]{\text{Accuracy/loss in 2D CNN in T1w, averaged  for  5  folds.}}
    \label{fig:t1w_lossplots}%
         \vspace{.1in}
\end{figure}

The training progress of this model is pictured in \text{Figure} \ref{fig:t1w_lossplots} in terms of training/validation accuracy and loss averaged over the five folds. Training resulted in a micro-averaged 81\% validation accuracy. \text{Table} \ref{tab:t1w_results} provides the accuracy (Acc), recall (TPR), precision (PPV), specificity (TNR), F1-score and MCC for this and the compared inferior 2.5D and 3D CNNs. We observe that in our setting, the 2D CNN model outperforms the more complex architectures. In small validation test datasets, we hypothesize that complex architectures like 3D CNN models are not an  appropriate option.

\begin{table}
\centering
\caption[Overall statistical results in T1w]{\text{Overall statistical results in T1w.}}
\label{tab:t1w_results}
\begin{tabular}{@{}lcccccc@{}}
\hline
\toprule
 & \multicolumn{6}{c}{\textbf{Overall Statistical Results/Metrics}} \\ \cmidrule(l){2-7} 
\textbf{CNN Model}  & \textbf{Acc} & \textbf{TPR} & \textbf{PPV} & \textbf{TNR} & \textbf{F1} & \textbf{MCC} \\ \midrule
2D CNN & 0.81 & 0.81 & 0.79 & 0.94 & 0.80 & 0.74 \\
2.5D CNN & 0.58 & 0.58 & 0.63 & 0.85 & 0.59 & 0.43 \\
3D CNN & 0.77 & 0.77 & 0.78 & 0.92 & 0.78 & 0.69 \\ \bottomrule
\end{tabular}\vspace{.25in}
\end{table}

\subsection{3D Fusion CNN versus 2D Fusion CNN in Brain Age Classification}
The early fusion of this data was to concatenate the three MRI sequences into the channel dimension of the input image. We trained our proposed fusion CNN models multiple times before selecting a set of hyperparameters that we used in all subsequent experiments. \text{Table} \ref{tab:fusion_results} shows the detailed metrics for this and the compared  3D in fusion models.The proposed 2D fusion model outperformed the 3D model at a 90\% accuracy across the cross-validated validation.

\begin{figure}
  \centering
  \includegraphics[width=1\linewidth]{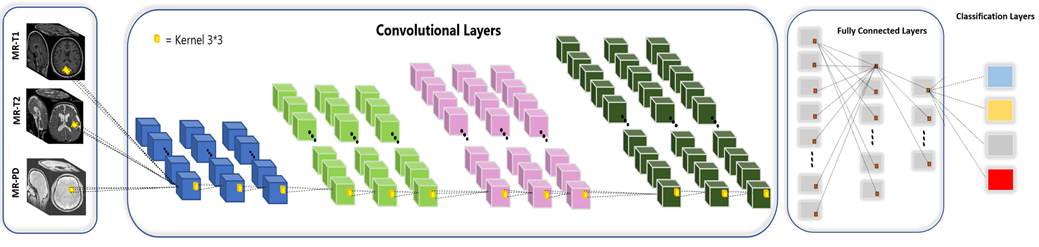}
  \caption{3D CNN architecture for early fusion.}
  \label{fig:fusion_model}%
\end{figure}

\text{Figure} \ref{fig:fusion_model} shows the 3D CNN architecture using the early fusion strategy on sets of 184 fusion inputs consisting of MR-T1w, T2w and PDw MRI sequences. Our approach to early fusion of this data is to concatenate the three MRI sequences into the channel dimension of the input image. Note that there is considerable motion between the contrast which we didn't account for through image registration. We trained our proposed fusion CNN models multiple times before selecting a set of hyperparameters (epochs to train, learning rate schedule, early stopping criteria etc.) that were used in all subsequent experiments. 

X <return>

\begin{figure}
\centering
     \includegraphics[width=1\linewidth]{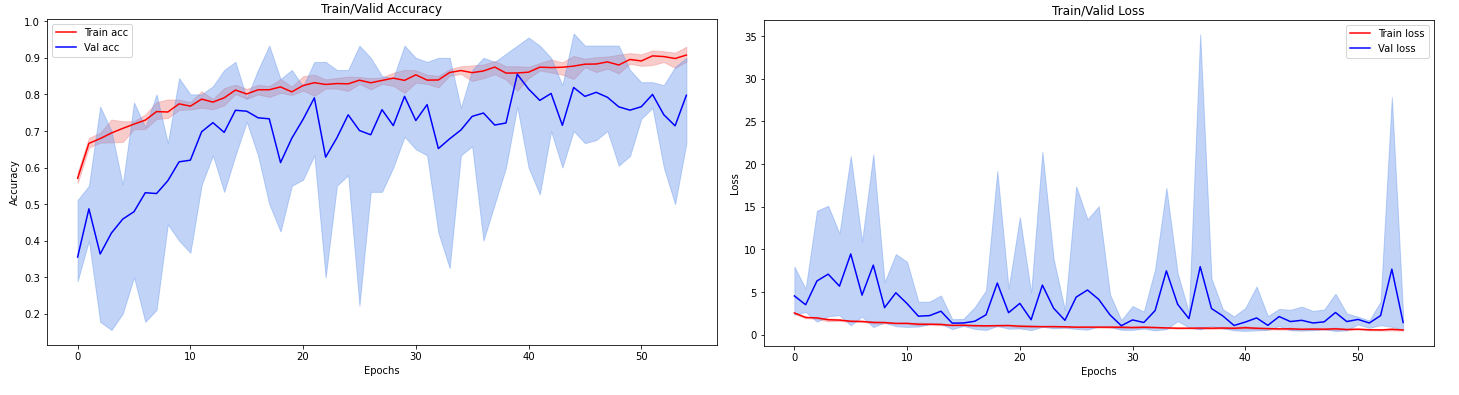}
     \caption[Accuracy/loss in 2D fusion CNN, averaged for 5 folds]{\text{Accuracy/loss in 2D fusion CNN, averaged for 5 folds.}}
     \label{fig:fusion_lossplots}%
        \vspace{.2in}
\end{figure}

\begin{figure}
\centering
\resizebox{\linewidth}{!}{
\begin{tabular}{c c}
\includegraphics[width=0.33\linewidth]{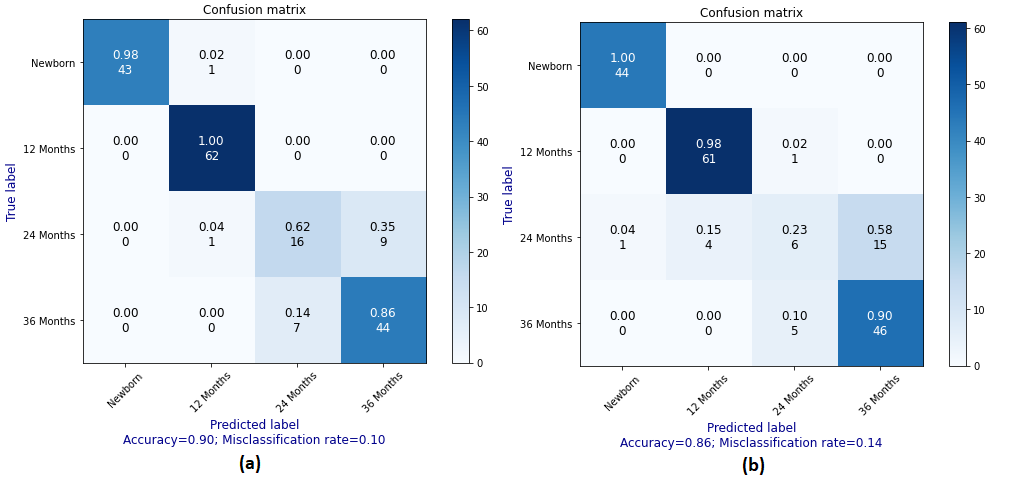}
\end{tabular}
}
  \caption[Confusion matrices for (a) 2D  
  and (b) 3D fusion CNN]{\text{Confusion matrices for (a) 2D  
  and (b) 3D fusion CNN.}}
  \label{fig:confusion_matrix}%
    \vspace{.1in}
\end{figure}
\begin{figure}
\centering
\includegraphics[width=.5\linewidth]{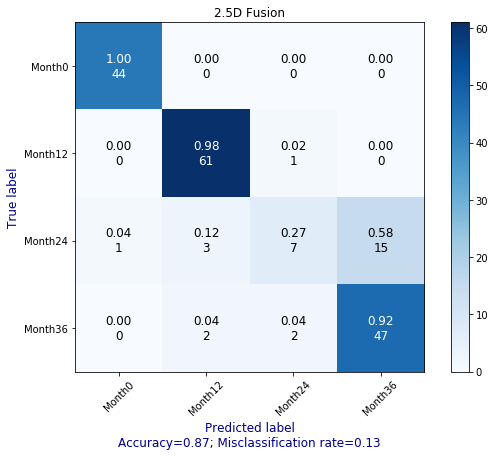}\\
    \caption[Confusion matrix for 2.5D fusion CNN]{\text{Confusion matrix for 2.5D fusion CNN.}}
  \label{fig:confusion}%
    \vspace{.1in}
\end{figure}

The averaged training/validation accuracy and loss of the 2D fusion CNN model is depicted in \text{Figure} \ref{fig:fusion_lossplots}, and \text{Table} \ref{tab:fusion_results} shows the detailed metrics for this and the compared 2.5D and 3D in fusion models, followed by the normalized confusion matrices in \text{Figure} \ref{fig:confusion_matrix} and \text{Figure} \ref{fig:confusion}. Consistent with the results on the T1w data alone, the proposed 2D fusion model outperformed the other architectures at a 90\% accuracy across the cross-validated validation. 

\begin{table}
\centering
\caption[Overall statistical results in the fusion CNN models]{\text{Overall statistical results in the fusion CNN models.}}
\label{tab:fusion_results}
\begin{tabular}{l c c c c c c c c c}
\hline
\toprule
 &  & \multicolumn{6}{c}{\textbf{Overall Statistical Results/Metrics}} &  \\ \cline{2-9} 
\textbf{CNN Model} & \textbf{Acc} & \textbf{TPR} & \textbf{PPV} & \textbf{TNR} & \textbf{AUROC} & \textbf{F1} & \textbf{MCC} & \textbf{Kappa 95\% CI} \\ \hline
2D CNN & 0.90 & 0.90 & 0.90 & 0.97 & 0.99 & 0.90 & 0.86 & (0.79-0.92)  \\
2.5D CNN & 0.87 & 0.87 & 0.86 & 0.92 & 0.98 & 0.87 & 0.83 & (0.74-0.90)\\
3D CNN & 0.86 & 0.86 & 0.85 & 0.91 & 0.98 & 0.86 & 0.83 & (0.75-0.89)\\ \bottomrule 
\end{tabular}\vspace{.25in}
\end{table}


We observed that using the fusion of common three MRI sequences improved performance in estimating brain developmental age. Despite the small dataset, the proposed simple 2D fusion CNN model achieved high accuracy in estimating brain developmental stage accurately. The fusion of multiple MRI sequences helped to extract more hierarchical features from the images. In infants, T1w, T2w and PD sequences were crucial to estimation of brain development.
We compared the overall statistical results of the proposed 2D and 3D CNN models using fusion of MRI sequences. While 3D CNNs are supposed to yield higher performance for analyzing volumetric data, we achieved better performance using the 2D fusion CNN model. We hypothesize that in cases where very little data is available, it is not possible to extract predictive information from the convolutional layers in the 3D model, since the variability of the ”normal” is too high to be captured in the limited dataset. Therefore, the use of the 2D model provides an advantage, although it is likely that in a scenario with more data the 2D model might be surpassed by a 3D model. CNN fusion models might be able to detect subtle micro- and macro-structural changes and aberrant connectivity variances that might go undetected at an age where early intervention could impactpatient care. In addition, these findings may be undetectable by the human eye until much later in the child’s development, preventing implementation of an early treatment plan. This paper illustrated that better performance could be obtained by passing the whole MRI volume from three MRI sequences into the 2D fusion CNN model.

\section{Summary of Results}

We compared the overall statistical results of the proposed 2D, 2.5D, and 3D CNN models using T1w and fusion of MRI sequences. While analysis of volumetric data using 3D CNNs would be expected to perform better, we achieved better performance using the 2D fusion CNN model. We hypothesize, also based on the additional 2.5D experiment, that in a setting with only few data, the extraction of predictive information in the convolutional layers is not possible in the 3D model, since the variability of the "normal" is too large to be captured in the limited data. Therefore, the 2D model has an advantage, though it is likely that in a scenario with small data, it might be surpassed by a 3D model. Further statistical results showed -- much in line with the clinical hypothesis -- that the fusion of multiple MR sequences improves the performance of age estimation in infants (\text{Table} \ref{tab:summary}). 

Stacking multiple MRI sequences might help to find high-level features associated with age from different MRI sequences. Therefore, CNN fusion models might be able to detect subtle micro- and macro-structural changes, as well as aberrant connectivity variances that may go undetected at an age where early intervention could have impacted patient care. In addition, these findings may be undetectable by the human eye until much later in the child's development, preventing implementation of an early treatment plan. This research showed better performance by passing in the whole MRI volume from three MRI sequences into the 2D fusion CNN model.


\begin{table}
\centering
\caption[Overall statistics in T1w/fusion MRI using 2D CNN model]{\text{Overall statistics in T1w/fusion MRI using 2D CNN model.}}
\label{tab:summary}
\begin{tabular}{l c c c c c c c c c}
\hline
\toprule
 & \multicolumn{7}{c}{\textbf{Overall Statistical Results/Metrics}} \\ \cline{2-8} 
\textbf{CNN Model} & \textbf{Acc} & \textbf{TPR} & \textbf{PPV} & \textbf{TNR} & \textbf{AUROC} & \textbf{F1} & \textbf{MCC} \\ \hline
2D CNN /T1w & 0.81 & 0.81 & 0.79 & 0.94 & 0.98 & 0.80 & 0.74 \\
2D Fusion CNN & 0.90 & 0.90 & 0.90 & 0.97 & 0.99 & 0.90 & 0.86 \\ \bottomrule 
\end{tabular}\vspace{.2in}
\end{table}


\section{Discussion}

Our counterintuitive finding was that a small model (in terms of depth and number of trainable parameters) outperformed more complex models with higher numbers of levels and parameters. Our result needs to be replicated on a larger, less redacted dataset in future research and the special value of working with 2D images needs to be elaborated upon with respect to voxel size and volume of tissue in the voxels. Ablation studies may provide exemplary methods for selecting optimal model complexity with regard to the classification task and the available dataset(s). The major goal of this work is to show that 2D CNN works better in the small number of samples. The original dataset has four age classes, even though the number of cases was too small specifically in 24 months to achieve reliable performances. Despite the small dataset, our method achieved high accuracy, and therefore suggests that further work is warranted to test whether BDAE using 2D fusion CNN might have a role in patient care.

Based on these results, 2D fusion CNNs could be used to determine the trajectory of normal brain development and neurodevelopmental age within the first three years of life. This approach could also prove useful of fusion multiple MR sequences in identifying otherwise undetectable abnormalities in brain development to improve performance.

\bibliography{report} 
\bibliographystyle{spiebib} 

\end{document}